\documentclass[12pt]{article}
\usepackage{graphicx,color}
\usepackage{apacite}
\usepackage{figcaps}
\usepackage{natbib}
\usepackage{amsmath}
\usepackage{amsfonts}
\usepackage{amssymb}
\usepackage[doublespacing]{setspace}
\usepackage{xspace}
\usepackage{fullpage}
\usepackage[ruled]{algorithm2e}
\newcommand{\Sc}{\emph{Saccharomyces cerevisiae}\xspace}
\newtheorem{thm}{Theorem}[section]

\printfigures

\begin{document}

\title{MCMC Inference for a Model with Sampling Bias: An Illustration using SAGE data.}

\maketitle

\begin{center}
{\bf Russell Zaretzki}  \\ \vskip .05in Department of Statistics, Operations, and Management Science\\The University of Tennessee\\
331 Stokely Management Center, Knoxville, TN, 37996  \\
\textit{email:} rzaretzk@utk.edu  \\ \vskip .1in \textbf{and} \\
{\bf Michael A.~Gilchrist}  \\ \vskip .05in Department of Ecology and Evolutionary Biology, \\The University of Tennessee\\
569 Dabney Hall, Knoxville, TN, 37996  \\
\vskip .1in\textit{email:} mikeg@utk.edu  \
\vskip .1in
\textbf{and} \\
\vskip .1in
\vskip .1in {\bf William M.~Briggs}  \\\vskip .05in Dept. of Mathematics   \\ Central Michigan University \\ \textit{email:} mattstat@gmail.com\\
\vskip .1in
\textbf{and} \\
{\bf Artin Armagan}  \\ \vskip .05in Department of Statistics, Operations, and Management Science\\The University of Tennessee\\
405B Aconda Court, Knoxville, TN, 37996  \\
\textit{email:} aarmagan@utk.edu
 \vskip .1in 
\end{center}

\begin{abstract}
This paper explores Bayesian inference for a biased sampling model in situations where the population of interest cannot be sampled directly, but rather through an indirect and inherently biased method.
Observations are viewed as being the result of a multinomial sampling process from a tagged population which is, in turn, a biased sample from the original population of interest. This paper presents several Gibbs Sampling techniques to estimate the joint posterior distribution of the original population based on the observed counts of the tagged population. These algorithms  efficiently sample from the joint posterior distribution of a very large multinomial parameter vector. Samples
from this method can be used to generate both joint and marginal posterior inferences. We also present an iterative optimization procedure based upon the conditional distributions of the
Gibbs Sampler which directly computes the mode of the posterior distribution. To illustrate our approach, we
apply it to a tagged population of messanger RNAs (mRNA) generated using a common high-throughput technique,
Serial Analysis of Gene Expression (SAGE). Inferences for the mRNA expression levels in the yeast \Sc are
reported.
\end{abstract}

%\begin{keywords}

\section{Introduction}
This paper develops methods for making Bayesian inferences about the composition of a population whose members
have different probabilities of being observed. Our approach applies to situations where the
categorical composition of a population is of interest and where some members of the population may be more
easily observed than others. The sampling process can be viewed as a multinomial process
where the probability of a sample being chosen will differ for each category in a known way. An example: survey samples of males and female birds that differ greatly in their coloration, markings, and
degree of vocalization.  Alternatively, sampling rates may be differentiated by age classes or species that
differ in their activity level or size.  Similar problems exist in studies of molecular biology where
observations are generally indirect and the ability to observe a molecule varies by type. Specific examples
include proteins that differ in their hydrophobicity or size. We will illustrate our ideas, by considering a
data set generated using Serial Analysis of Gene Expression (SAGE), a bioinformatic technique used to measure mRNA expression levels.

SAGE is a high-throughput method for inferring mRNA expression levels from an experimentally generated set of
sequence tags. A SAGE dataset consists of a list of counts for the number of tags that can be unambiguously
attributed to the mRNA of a specific gene. These observed tag counts can be thought of as a sample from a much
larger pool of mRNA tags. The tag counts are then used to make inferences about the proportion of mRNA from each
gene within the mRNA population from a group of cells.  The standard approach for interpreting SAGE data is through the use of a multinomial sampling model
\citep{Velculescu1995,Velculescu1997}. \citet{Morris2003} directly applied a Bayesian
multinomial-Dirichlet model to the observed vector of tag counts. This approach improves upon most earlier work
by considering simultaneous inference on all proportions. They provide a simple computationally tractable
approach and consider the result of the statistical shrinkage effect which offers improved estimates for
proportions with low tag counts while underestimating the expression proportions for tags with large counts.
This leads them to propose a mixture Dirichlet prior in order to mitigate the propensity to underestimate highly
expressed genes.

An alternative analysis, which was not based on a multinomial model directly, was developed by
\cite{Thygesen2006}, who modeled the marginal distribution of the counts across tag types as
though they were independent observations from a Poisson distribution. They applied a hierarchical
zero-truncated Poisson model with mean parameter which followed either a gamma or log-normal distribution. A
non-parametric adjustment factor was required in order to correctly capture the overabundance of larger counts.
Similar analysis \citep{Kuznetsov2002,Kuznetsov2006} modeled SAGE counts using a discrete Pareto-like
distribution.  These studies found that this model
effectively predicted counts greater than zero. One drawback is that the variance of
expression cannot be explicitly separated from sample variance.  However, in the context of differential
expression, \cite{Baggerly2003} suggests that treating genes individually offers less power than a model, such
as the multinomial, which incorporates all tags simultaneously.

One common thread is the explicit assumption that an mRNA's frequency
in the sampled tag pool is equivalent to its frequency in the mRNA population from which the tag pool was
derived. However, because the ability to form tags from an mRNA transcript varies from gene to gene, the tag
pool that is sampled is actually a biased sample of the mRNA population \citep{Gilchrist2007}.  \cite{Gilchrist2007} illustrated how the tag formation bias could be estimated and incorporated through the
calculation of a gene specific tag formation probability. The complicating factor is that the sampling bias of
one gene is not only a function of its own tag formation probability, but is also inversely proportional to a
mean tag formation probability where the contribution of each gene to this mean is weighted by its frequency in
the mRNA population, i.e.~the very parameters we wish to estimate.

As a result of the sampling bias, the probability of observing an individual gene depends on both its tag
formation probability and the distribution of these probabilities across all other genes weighted by their mRNA
frequency. \cite{Gilchrist2007} derive an implicit solution for the maximum likelihood and joint posterior mode
estimators of the composition of the mRNA population. However, there are a number of numerical stability issues
that severely limit the range of prior parameters that can be used, i.e. those priors with appreciable weight relative to the observed sample sizes. There is also the restrictive assumption that the contribution of an individual gene to the weighted average is small in order to derive approximations
for the marginal credible sets for a given gene.  We introduce several new hierarchical models for the posterior distribution of the mRNA proportions given the bias in the observed data. These methods are more robust, flexible, and require fewer assumptions than the analytic approach outlined in \cite{Gilchrist2007}.

In addition to the bias introduced through the tag formation process itself, other steps
in the experimental process can introduce errors which will affect the ability to interpret SAGE data. These
include sampling error, sequencing error, non-uniqueness and non-randomness of tag sequences
\citep{Stollberg2000}. For example, the use of PCR to amplify mRNA samples can introduce copying errors into tag
sequences. Tags are identified via DNA sequencing, which is an imperfect processes and errors occur on a per
nucleotide basis. As a result the error rate depends on the length of the tag's generated. A number of
sophisticated techniques have been developed to correct for such errors
\citep{ColingeAndFeger01,AkmaevAndWang04,BeissbarthEtAl04}. In this study we ignore such
complications but believe that it is also possible to address these sources of error by incorporating them into
a more complicated model of the experimental process.

Because we are using SAGE data as an illustration, Section \ref{SAGE} describes the experimental process used to
generate a SAGE data set as well as the sources of biases that can result and additional sources of errors.
Section \ref{Statistics} discusses the basic sampling and inference. Section \ref{Gilchrist} introduces
a model for tag formation, and discusses the variation in tag formation. It also introduces a posterior
distribution which explicitly considers bias in tag formation and its consequences on inferences for expression
rates. Section \ref{MCMCmodels} introduces several approaches to simulate and estimate the posterior
distribution. These include both MCMC based sampling methods as well as direct optimization of the posterior.
Finally, Section \ref{Conclusion} discusses implications of this method for analysis of SAGE and compares this
methodology to that discussed by \cite{Morris2003}.  Further applications of this methodology to SAGE along with
more general applications of the models are also discussed.

\section{The SAGE Methodology \label{SAGE}}
The goal of a SAGE experiment is to sample the mRNA population within a group of cells and was developed by \cite{Velculescu1995}. Broadly speaking, the SAGE method generates a set of short sequence-based cDNA tags from the mRNA population of
a group of cells. Initially, a pool of mRNA is extracted from a group of cells. The unstable single stranded
mRNA is reverse transcribed into a double stranded DNA copy (cDNA) using a modified primer that allows the cDNA
to be bound to a streptavidin bead. The cDNA is cut into small `tags', using
restriction enzymes, whihc are concatenated together into longer cDNA molecules referred to as
concatemers. These concatemers are then amplified and sequenced. A cleavage motif of the anchoring enzyme
allows the start and stop points of tags to be identified in the sequence data. Each time a tag from a specific
gene is observed in the sequence data, it contributes a count to the dataset.  The data is then summarized by the number of tag counts for each individual gene within a genome, whihc are then used to make inferences about the mRNA population within a group of cells.

%There are a number of systematic problems that result from the SAGE experimental process which affect the nature and strength of inferences which can be made. We discuss three problems below: tagging bias, tag ambiguity, and tagging errors, noting that our technique only addresses the first.

The restriction enzymes used to create the tags only cut at very specific sites within
the cDNA. For example, the restriction enzyme used by \cite{Velculescu1995} could only cleave the cDNA at the
four nucleotide motif CATG. Thus, tags can only be created at specific points within a gene. Because the site
where the cDNA is cut serves as an `anchor' between tags, the restriction enzyme used is often referred to as
the anchoring enzyme (AE). We also refer to the specific points in the cDNA cleaved by the anchoring enzyme as AE sites.

\begin{figure}[htbp]
    \caption{ Plot showing cDNA cleavage sites for SAGE with associated probabilities of tag formation.\label{cDNAfig}}
    \begin{center}
        \scalebox{.9}[.9]{\includegraphics{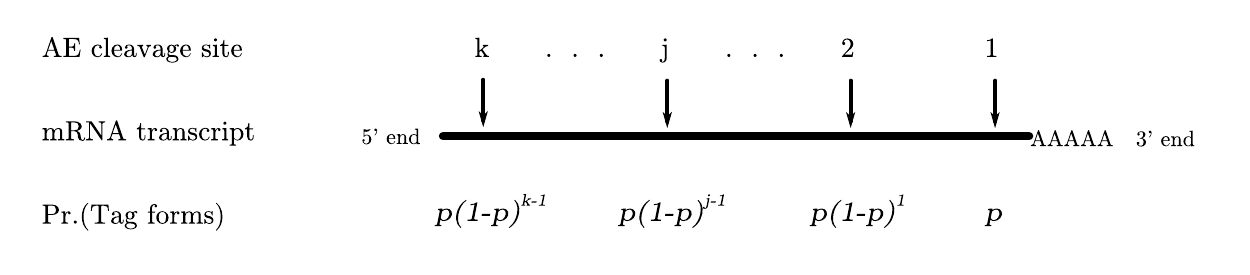}}
    \end{center}
\end{figure}

While some genes will have no AE sites, many genes will have multiple sites at
which the AE could cleave the cDNA. Given that the AE is expected to act in a site-independent manner,  a single
cDNA molecule can be cut by the AE in multiple places. However, because only the fragment of cDNA that is
attached to a streptavidin bead is retained during the experimental process,  the site closest to the bead
(i.e.~the 3' most site) that is actually cleaved is the only site that can lead to an observable tag (see Figure
\ref{cDNAfig}). If the AE worked with 100\% efficiency, then each mRNA could only lead to one observable tag.
However, the cutting efficacy of the AE is always less than 100\% and as a result multiple tags may be generated
due to the multiple copies of a gene's mRNA in the mRNA population of interest. We  discuss the overall
probability that a gene's mRNA transcript can be tagged as well as the expected distribution of the tags it can
form in Section \ref{Gilchrist}. The critical point is that differences in the number of AE sites between genes result in different probabilities that their mRNA will form a tag. Genes whose mRNA
transcript lacks any AE site represent a most extreme example of this bias since such genes have have zero
probability of forming a tag. Such genes are impossible to observe using the SAGE methodology.
A further complication is a lack of independence between the mRNA sequences of different genes. This is mainly due to the fact that most genes are the result of gene duplication events. While most
genes do contain an AE site, many of these sites lead to non-unique or `ambiguous' tags which cannot be readily
assigned to a particular gene. Ambiguous tags are `uninformative' using current technologies; tag-to-gene
mapping is discussed at length in \citet{Vencio2006}.

Experimentalists have attempted to deal with the problem of ambiguous tags by increasing the size of the tag
and, thus, decreasing the probability a tag can be attributed to more than one gene. The actual tag length
depends on the specific tagging enzyme used (which is another type of restriction enzyme), but is invariant
within an experiment. Initially SAGE tags were 14 bp long. However, four of these bp reflect the cutting motif
of the AE and, as a result, were shared by all tags. Experimental advances have been able to extend the tag
length to over 20 bp.
For example, `SuperSAGE' techniques can lead to tags up to 26 bp long \citep{MatsumuraEtAl03,MatsumuraEtAl06}.
Unfortunately, extending the length of a tag comes at a cost \citep{Stollberg2000}. This is
because neither the reverse transcription, the PCR amplification, or sequencing process is error free. For
example, \cite{Velculescu1997} estimate that the sequencing error rate of a tag is $0.007/ \text{bp}$, thus the
probability of obtaining an error free tag decreases geometrically with sequence length, ranging from
approximately 10\% to 15\% in 14 and 22 bp tags, respectively. Transcription errors, either by the cell or
during the conversion to cDNA or amplification by the experimentalist, can create either novel `orphan' tags
which cannot be mapped to a particular gene or misleading tags which are attributed to the wrong gene.

\section{Basic Statistical Inference \label{Statistics}}
Before the SAGE tag counts are considered, we assume that the data is processed to retain only informative tags, i.e.~all ambiguous or orphan tags are removed, as is  standard practice. The result can be viewed as a vector of observed tag counts for individual genes, i.e.~${\bf T} = (t_1, \ldots t_k)$
where $k$ is the number of genes that contain at least one unambiguous AE site and $t_i$ is the sum of all
counts which can be uniquely attributed to gene $i$. It is natural here to view
this vector as a sample from a multinomial distribution with $k$ categories. Thus $\mathbf{T} \sim \text{Multi}\left(T_{tot},\boldsymbol{\theta}\right)$ where $\theta_i$ represents the frequency of gene $i$ tags
in the tag pool and $T_{tot}$ represents the total number of informative tags sequenced, i.e.~$T_{tot} =
\sum_{j}t_j$. Until recently, inferences about a gene's frequency in the mRNA population, the population of
interest, were assumed to be equivalent to any inferences made about its tag frequency, i.e.~ the tag pool was
viewed as an unbiased sample of the mRNA population. As pointed out by \cite{Gilchrist2007}, this equivalence
only holds when all genes have the same tag formation probability. Given the variation between genes, this
condition will never be met. Consequently,  genes with greater than average tag formation probabilities will be
over-represented in the tag pool. Conversely, tags with a lower than average tag formation probability will be
under-represented. This sampling bias in the formation of the tag pool, however, can be estimated and used to
correct the inferences made from the SAGE data. The first step in adjusting for this bias is the calculation of
the tag formation probabilities for every gene in the genome.

\subsection{Biased Sampling and Tag Formation Probabilities \label{Gilchrist}}
As pointed out in Section \ref{SAGE}, only cDNA that is attached to a streptavidin bead is retained during
the experimental process. As a result, tags are created from the 3' most AE site that is actually cleaved
(Figure \ref{cDNAfig}). Let $k_i$ be the number of restriction enzyme sites which may be cleaved by the AE on
mRNA generated from gene $i$, let $p$ be the probability the AE will cleave a site, and assume that cleavage
probability is independent between sites and does not vary between genes. If we label sites 1 through $k_i$
starting at the 3' most site (i.e. the site closest to the bead; see Figure \ref{cDNAfig}). It follows that the
probability of generating a tag through AE cleavage at site $j \in \{1, \ldots, k_i\}$ is $(1-p_i)^{j-1}p_i$.
This corresponds to the probability of no cleaving in sites 1 to $j-1$, and a cleaving at site $j$. Note
that this probability is independent of what happens at the AE sites 5' from the $j$th site. Thus, the
probability of creating a tag at site $j$ on an mRNA generated from gene $i$ follows a geometric distribution.
The fact that the expected distribution of tags varies with AE cleavage probability $p$ can be used to estimate
$p$ from the set of intra-genic tag distributions \citep{Gilchrist2007}.

If the set of tags which can be created from the AE sites within a gene's mRNA trascript are all informative
(i.e.~can be uniquely assigned to a single gene),  the probability of generating any of the possible tags from
the $i$th category of mRNA is
\begin{equation}
\label{eq:phisum}
\phi_i = \sum_{j=1}^{k_i} (1-p)^{j-1}p = 1-(1-p)^{k_i}.
\end{equation}
When ambiguous tagging sites occur within a gene's mRNA transcript, say sites $l$ and $m$, these sites are
simply excluded from the summation over $j$ in eq.~(\ref{eq:phisum}). The reduction in $\phi_i$ due to tag
ambiguity is greatest when an ambiguous tag is formed at the 3' most AE site (site 1 in Figure \ref{cDNAfig})
and is least when such a tag occurs in the 5' most AE site (site $k_i$). But note that it is possible for
transcripts to have multiple ambiguous AE tag formation sites. For purposes of our analysis, we will assume that
the value of $\phi_i$ has been estimated for all genes and treat these as {\it known} constants (see
\cite{Gilchrist2007} for a more details on these calculations).  The variation in the tag formation probability between genes stems from the two basic facts. First,  genes vary
in the number of AE sites they contain. Second, genes also vary in the number and location of AE sites which are
ambiguous. As a result the tag pool represents a biased sample of the mRNA population of interest. The degree of
bias exhibited is a function of both the distribution of tag formation probabilities across the genome, and the
distribution of these genes within the mRNA population itself, as we will now demonstrate.

\subsection{Maximum Likelihood and the Joint Posterior Distribution\label{model}}
As discussed earlier, an obvious model for the observed tag data is the multinomial distribution.
Let $T_{tot} = \sum_{i=1}^k t_i$ be the total count of all observed tags.
Then
\begin{equation*}
P(T|\theta) = {T_{tot} \choose t_1,t_2, \ldots, t_k} \prod_i \theta_i^{t_i}
\end{equation*}
where $\theta_i$ are the frequencies of tags in the tag pool.
Section \ref{Gilchrist} indicates that frequency of SAGE tags generated from gene $i$ is,
\begin{equation}
 \theta_i = \frac{m_i \phi_i}{\sum_j m_j \phi_j}\label{eq:theta.def}
\end{equation}
where, we remind the reader, that $\phi_i$ is {\it known} and represents the probability that a mRNA strand from gene $i$ in the
group of cells will be converted into a tag and where $m_i$, the quantity of interest, is the relative
proportion of mRNA from gene $i$ in those cells. Hence, the bias corrected likelihood model is,
\begin{equation}
P(T|\mathbf{m}) = {T_{tot} \choose t_1,t_2, \ldots, t_k} \prod_i \left(\frac{m_i \phi_i}{\sum_j m_j \phi_j}
\right)^{t_j} \label{eq:mlike}
\end{equation}
Because the vector $\mathbf{m} = \{m_1, m_2, \ldots m_k\}$ is a vector representing relative proportions, or
equivalently the probability that an individual mRNA strand represents gene $i$, the components of $\mathbf{m}$
must satisfy the conditions: $m_i \geq 0$ and $\sum_{i=1}^k m_i = 1$.

Direct maximization of the likelihood with respect to the parameters $m_i$ is straightforward via the invariance
property of the MLE. Consider the observed sample proportion $\hat{\theta}_i = t_i/T_{tot}$. Given constraints
on the possible values of $m_i$ and equation (\ref{eq:theta.def}),  the MLE estimates of $\hat{\mathbf{m}} =
\{\hat{m}_1, \hat{m}_2, \ldots, \hat{m}_k\}$ must satisfy the following equality,
\begin{equation}
\label{eq:implicit.mhat} \sum_i^k \frac{\hat{\theta}_i}{\phi_i} = \frac{1}{\sum_{i}^k \hat{m}_i \phi_i}.
\end{equation}
While estimation of the MLE is relatively straightforward, estimating confidence intervals for $\mathbf{m}$ is
difficult. The existence of the normalization term $\sum_j m_j \phi_j$ in the denominator of the l.h.s.~of eq.
(\ref{eq:implicit.mhat}) makes computation of the information matrix taxing, particularly for the large
dimensional vectors encountered when working with SAGE data (i.e. $k$ is of the order $10^3$ to $10^4$). In
addition, because the number of categories is generally within an order of magnitude of, or possibly even
greater than, the number of tags sampled, most categories have either zero or only a few observations. These
relatively small cell counts make any inferences based on asymptotic approximations questionable.

As an alternative, we consider a Bayesian approach to the problem. The constraints lead naturally to the
assumption of a Dirichlet prior on $\mathbf{m}$, \[ \mathbf{m} \sim D_k(\alpha_1,\ldots, \alpha_k) =
\frac{\Gamma(\sum \alpha_i)}{\prod_{i=1}^k \Gamma(\alpha_i)} \prod_{i=1}^k m_i^{\alpha_i -1}. \] Combining this
prior distribution with the likelihood function Equation \ref{eq:mlike} leads to a posterior distribution
proportional to the product
\begin{equation}
[\mathbf{m}|\mathbf{T}] \propto  (\sum_{i=1}^k m_i \phi_i)^{T_t} \prod_{i=1}^k (m_i \phi_i)^{\alpha_i +
t_i-1} \label{eq:postdist}
\end{equation}
\cite{Gilchrist2007} discuss methods for the direct maximization of this quantity and also discuss the choice of a prior and its consequences on the marginal inferences of $m_i$ for particular values of $i$.
%The un-normalized log-posterior can now also be easily derived as
%\begin{equation*}
%\log \pi(\mathbf{m}|\mathbf{T},\phi) \propto T_{tot} \log(\sum_{i=1}^k m_i \phi_i) + \sum_{i=1}^k (\alpha_i + t_i - 1) \log(\alpha_i).
%\end{equation*}
A main inferential difficulty is the existence of the normalizing term $\sum_{i=1}^k m_i \phi_i$ which is
required to force the values $\theta_i$ to sum to one. Because the form of the posterior is not standard, this leads to difficulties when trying to numerically estimate the normalizing constant.
Therefore, we adopt a Monte-Carlo approach to inference on the posterior distribution.

%
%{\it DO I NEED TO SAY SOMETHING ABOUT A JACOBIAN HERE OR THE LACK THEREOF: EVEN IF ONLY TO MAKE SURE EVERYONE
%HAS THOUGHT ABOUT THIS.  SURE (mike).

\section{Gibbs Sampling Strategies for Posterior Analysis \label{MCMCmodels}}
We explicitly consider three strategies to simulate from the posterior distribution given by eq.
(\ref{eq:postdist}). The first strategy is presented in Subsection \ref{BinPois}. This approach extends the
Binomial - Poisson hierarchical model of Cassella and George given in \cite{Arnold1993}. Given a random count
$T_i$, this model proposes a binomial distribution $T_i \sim Binomial(g_i,\phi_i)$. $g_i$, the number of trials,
is assumed to be Poisson distributed and represents the total number of mRNA's of a given type in the cells.
Here, we extend the model to take into account the multinomial nature of the original mRNA population. The
second strategy is presented in Subsection \ref{BinMulti}. This approach proposes that a count $T_i \sim
Binomial(g_i,p)$ and that the vector $\mathbf{g} \sim Multinomial(N,m)$. We believe this strategy  represents
the data generating process of SAGE slightly more realistically than the first strategy. Finally, in Subsection
\ref{Conjugate} we consider a missing data strategy. This approach applies a modified version of the conjugate
Dirichlet-Multinomial model and is more computationally expedient than the first two strategies. Overall, each
of the three strategies offers its own unique strengths and weaknesses, which we discuss below.

The algorithms discussed were all applied to a published SAGE data set which analyzed the yeast \Sc
\citep{Velculescu1997}. In particular, data collected from the log-phase were analyzed. There were 6096 genes
included in the data. The maximum tag count was 392 and the minimum was zero.  3560 of the genes included had
observed counts of 0. Gene dependent sampling probabilities ranges from $\approx .003 \ \mbox{to} \ 1$. Tag
assignments to unique genes and assignment of non-unique tags for this analysis were described in
\cite{Gilchrist2007}. All simulations described below were computed using $\mathtt{R}$ (Version 2.3.1) on
dual core Intel desktop computer running Linux Fedora 5.

In what follows, the symbol $\mathbf{T}$ will represent the vector of observed tag counts, $\mathbf{\phi}$ the known
vector of gene dependent sampling probabilities, $\mathbf{g}$ a latent vector representing the actual number of
mRNA's from each gene, $\mathbf{m}$ the vector of mRNA proportions and $N$ the total mRNA copy number in the
cell.  Throughout this analysis, we assume that gene dependent tag formation probabilities $\phi_i$ are known
constants.

%[Data Set Used,
%Number of Categories,
%Choices of parameters.
%Time it took to run.
%Machine Used.
%Results and Confidence Intervals for a range of genes.]

\subsection{Gibbs Sampling based on a Dirichlet-Poisson-Binomial Model.} \label{BinPois}
%Code is Gibbs1Unix1 and Gibbs1Unix,

Mechanically, the SAGE technique proceeds as discussed in Section \ref{SAGE} where the pool of mRNA from the
cells are first converted to cDNA, tagged and then amplified. In the model discussed below, inspired by Cassella
and George, we assume that a fixed population of mRNA's exists of size $N$ which we will suppose is random
within the cells. The total size $N$ is determined by a gamma distribution which is rounded off to the nearest
integer. The choice of gamma here is convenient due to its role as a conjugate prior. Given the population of
size $N$, the relative proportions $m_i$ of different categories of mRNA are unknown and may be modeled as a
Dirichlet distribution with prior $\mathbf{\alpha} = (\alpha_1, \ldots , \alpha_k)$ where the $\alpha_i$ will often be chosen
to be identical, i.e. $\alpha_i = c \ \forall i$ and some constant $c$.

Because cells contain mRNA transcripts from thousands of different genes, the probability of seeing any
particular gene is low. It is therefore logical to assume, given $N$ and $m_i$, that the actual number of mRNA's
of a certain type $g_i, i = 1, \ldots k$, extracted from the group of cells, satisfies a Poisson law with mean
$\mu = N \cdot m_i$. Finally, the restriction enzyme process generates a tag count $t_i$ for a particular
complimentary DNA strand in a binomial fashion based upon the tag formation probability $\phi_i$.  Hence, the hierarchical data generating mechanism follows, $T_i \sim  Binomial(g_i,\phi_i)$, $g_i  \sim Poisson(N*m_i)$, $\mathbf{m}  \sim  Dirichlet(\alpha_1, \ldots , \alpha_k)$,  $N \sim ceiling(Gamma(\gamma_1,\gamma_2))$.  We refer to this model as the Dirichlet-Poisson-Binomial (DPB) model. A first weakness of this model is that the
sum of all counts $\sum g_i$ may not add up to the total counts $N$. However, this approach allows us to infer
the total population size $N$. In constructing this hierarchical model, we have essentially provided a
multivariate extension of the model investigated in \cite{Thygesen2006} while integrating the sampling bias.  Together, the elements described above give a joint posterior distribution,
\begin{eqnarray}
[{\mathbf{m}},{\mathbf{g}},N|\mathbf{T},\boldsymbol{\alpha},\gamma_1,\gamma_2] &\propto& \prod_{i=1}^{k} {g_i \choose t_i} \phi_i^{t_i}(1-\phi_i)^{g_i-t_i} \frac{e^{-Nm_i}(N m_i)^{g_i}}{g_i !} \nonumber \\
&&\times {m_i}^{\alpha_i-1} e^{-\gamma_2 N} N^{\gamma_1-1} \nonumber \end{eqnarray}
From this expression, we can deduce the set of full conditional distributions \citep{Ghosh2006},

\begin{eqnarray}
\left[m_i|t_i,g_i,N,\boldsymbol{\alpha},\gamma_1,\gamma_2 \right] &\sim & \Gamma(g_i+\alpha_i,N)\nonumber \\
\left[g_i|\mathbf{T},\mathbf{m},N,\boldsymbol{\alpha},\gamma_1,\gamma_2 \right] & \sim & t_i + Poisson(N m_i(1-\phi_i)) \nonumber \\
\left[N|\mathbf{T},\mathbf{g},\mathbf{m},\boldsymbol{\alpha},\gamma_1,\gamma_2 \right] &\sim &
\Gamma(\sum_{i=1}^k g_i +\gamma_1,1+\gamma_2) \nonumber \end{eqnarray}

A Gibbs sampler can be implemented based on this full set of conditionals; however, a second weakness of this
approach is the need to update each value of $m_i$ independently. Because the vector ${\bf m}$ is typically on
the order of thousands or tens of thousands, this is a costly computation.

In order to minimize autocorrelation between samples, our experiments stored only the last of every hundred
samples for analysis. Inference for means was based upon the final 1000 of these stored samples drawn after an
extensive burn in period of 400,000 simulations. The two prior parameter vectors tested were $
\boldsymbol{\alpha} =1 \ \mbox{ and } \alpha_i=1/k \ \forall i$.  $k = 6096$ is the number of genes in the
sample. The distribution of population size, $N$, used a shape parameter $\gamma_1 = 100$ and prior scale
$1/\gamma_2 = 200$. The mean of a gamma with these parameters is 20000 which is somewhat larger than a natural
estimate of the population size, $\sum_i( t_i/\phi_i) = 16438.81$. This was chosen intentionally to determine if
the posterior would converge to a reasonable estimate of N.  Experimentalists assume that there are $\sim
15,000$ mRNA in a cell which also justifies this assumption.

As mentioned above, in addition to lags of 100 between samples, a very large burn-in period was used in the
analyses of Sections \ref{BinPois} $-$ \ref{Conjugate}. In order to ensure that autocorrelation did not
adversely effect parameter estimates, some basic convergence analysis was performed.  Convergence was evaluated
on 2 sampled quantities, the total mRNA population size $N$ and the gene at the third locus in the data set:
YAL003W.  Both quantities have direct inferential value and autocorrelation could adversely affect estimates.
YAL003W was selected randomly among the set of genes with medium to large tag counts. Out of the 12799 tags
observed, 32 could be attributed to gene YAL003W.  Direct autocorrelations of both sequences were tested at
various lags.  Table \ref{DPBcor} examines correlations among samples for different lags and suggests that for
inference on proportions $m_i$, autocorrelations are very low between samples with lags as small as 10.
Conversely, correlations, for sample size $N$ are extremely high leading to much less reliable inference.

In practice, if only the vector $\mathbf{m}$ is of interest, far fewer samples are needed which would
significantly speed up the algorithm. Physical timings for runs of these and other algorithms are given in Table
\ref{timing}. These show that as executed, this algorithm required about 11-12 hours to run.
If only every 10th sample were collected, 50,000 total samples, the algorithm would have required approximately 1 hour.

\begin{figure}[htbp]
\begin{center}
\caption{\label{fig1} Probability estimates and inferences for the Dirichlet -Poisson -Binomial with each
$\alpha_i=1$. The 20 genes with the largest tag counts are arranged in decreasing rank order along the X-axis.
The observed tag proportions are marked in dark circles, the standard MLE in dark triangles. The analytically
computed posterior mode $\alpha_i=1$ coincides exactly with the MLE.  Also included are the estimated posterior
mean and upper and lower 95\% Bayesian Credible bounds based on MCMC sampling.}
%\figbox*{}{}{\includegraphics[scale=.75]{figDPB1.pdf}}
\includegraphics[scale=.75]{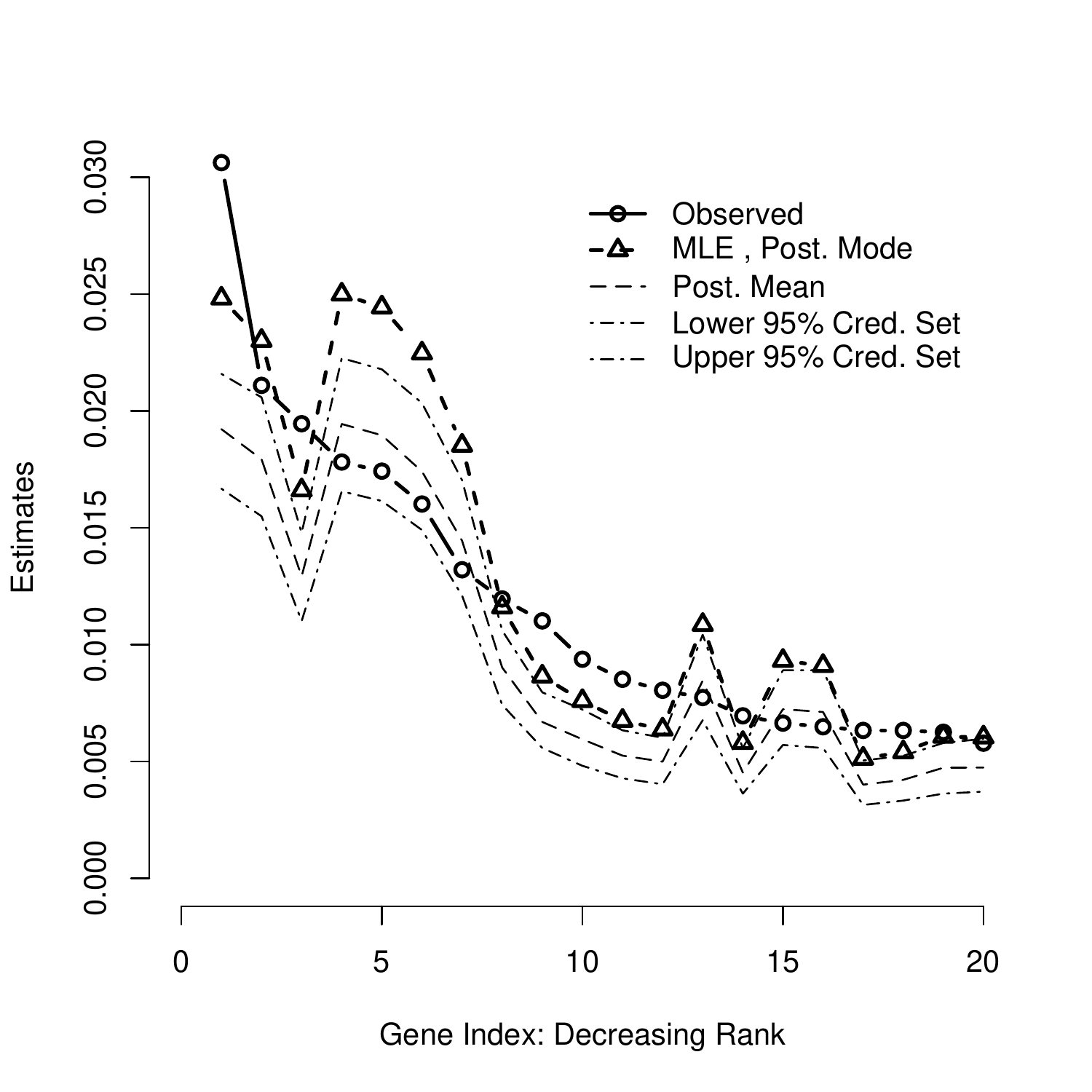}
\end{center}
\end{figure}

%In addition, the $\mathtt{R}$ function $\mathtt{varest}$ found in \cite{Gray2003} was used to estimate standard
%errors for means of a correlated sequence by using variation in means of blocks of size m. It assumes adjacent
%blocks are correlated, but blocks at greater lags are not.  The resulting correlated and uncorrelated standard
%errors are reported in Tables \ref{SEM} and\ref{SEN}.  Again, these plots show very little effect of correlation
%on the estimated standard errors.
In order to understand the behavior of the proportions, we compared various estimates of the $m_i$. These include
posterior means from the Gibbs sampler, standard MLE's, corrected MLE's based on Formula \ref{eq:implicit.mhat},
and analytically derived posterior modes from \cite{Gilchrist2007}. In addition, 95\% posterior credible regions
are also computed. Figure \ref{fig1} plots the above quantities versus rank for the 20 genes with largest tag
counts. Rank 1 corresponds to the most frequent gene and the prior used is $\boldsymbol{\alpha}=1$ . As one
would expect, the corrected MLE and the posterior mode coincide perfectly in this case.  Both posterior means
and posterior modes based on the biased sampling correction deviate strongly from the standard MLE for many of
the most frequently tagged genes.  Posterior modes and means follow nearly identical trends with the modal
estimates being uniformly larger. As one can see, modal estimates typically lie slightly above the $97.5^{th}$
percentile of the simulated posterior distribution in this case.

\begin{figure}[htbp]
\begin{center}
\caption{\label{fig2} Probability estimates and inferences for the Dirichlet -Poisson -Binomial with each
$\alpha_i=1/k$. The 20 genes with the largest tag counts are arranged in decreasing rank order along the X-axis.
The observed tag proportions are marked in dark circles, the standard MLE in dark triangles. In this case
analytically derived posterior modes deviate substantially from MLE.  However, estimated posterior mean is
identical to MLE in this case.  Upper and lower 95\% Bayesian Credible bounds are also given.}
\includegraphics[scale=.75]{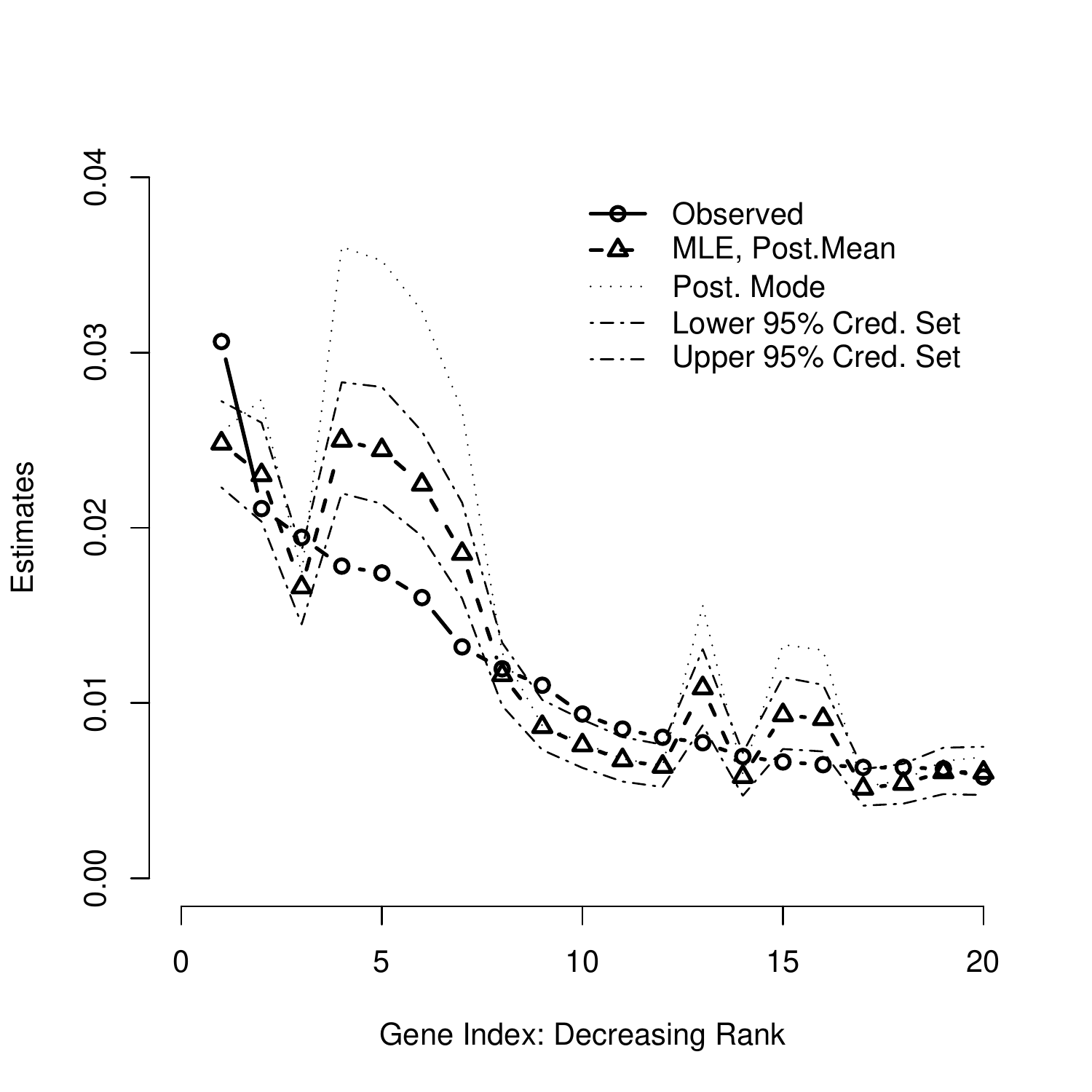}
\end{center}
\end{figure}

Figure \ref{fig2} summarizes results when each $\alpha_i = 1/k$.  This parametrization corresponds to a
U-shaped prior distribution.  Here one can see the $i^{th}$ corrected MLE and posterior mean now coincide almost
perfectly while the posterior mode lies at or above the $97.5^{th}$ percentile of posterior samples.  Again, all
estimators deviate significantly from the unadjusted MLE's.  Due to the similarity of the corrected MLE and
posterior mean, it is tempting to interpret the posterior intervals as frequentist confidence intervals.  To
date, no work has been done to verify the frequentist coverage properties of these intervals.

\begin{table}
\caption{Autocorrelation Estimates for the Dirichlet-Poisson-Binomial (DPB), Dirichlet-Multinomial-Binomial (DMB), and Missing Data algorithms. \label{DPBcor}}
\begin{center}
\begin{tabular}{ c rrrrrr }
  \hline
        &\multicolumn{2}{c}{DPB} & \multicolumn{2}{c}{MB} & \multicolumn{2}{c}{MD}\\ \hline
    & N &  N &  N & N & $r$ & N \\
   Lag & {\small ($\alpha = 1$)} &  {\small ($\alpha = 1/k$)} &  {\small ($\alpha = 1$)} & {\small ($\alpha = 1/k$)} & {\small ($\alpha = 1$)} & {\small ($\alpha = 1/k$)} \\ \hline
    10 & 0.905 &  0.024 & 0.0001  &-0.011 &  0.606 &  0.007\\
    20 & 0.863 &  0.018 & 0.020   & 0.020 &  0.344 & -0.012\\
    40 & 0.783 & -0.029 & -0.027  &-0.010 &  0.088 & -0.034\\
    80 & 0.666 &  0.027 & -0.009  &-0.018 & -0.011 &  0.011\\
  \hline \end{tabular}
\end{center}
\end{table}

\begin{table}
\caption{\label{timing} Execution Times for Algorithms. 5 million samples collected. Number of genes k = 6096. Algorithms compared are Dirichlet-Poisson-Binomial (DPB) of Section \ref{BinPois}, Dirichlet $-$ Multinomial Binomial (DMB) of Section \ref{BinMulti} and the Missing Data method (MD) described in Section \ref{Conjugate}.}
\begin{center}
\begin{tabular}{ c c r }
  \hline
  % after \\: \hline or \cline{col1-col2} \cline{col3-col4} ...
  Method & $\alpha$ & run time (min.)\\
  DPB &  1  & 750\\
  DPB & 1/k & 730\\
  DMB & 1   & 30\\
  DMB&  1/k & 50\\
  MD &   1  & 20\\
  MD &  1/k & 19\\
\hline \end{tabular}
\end{center}
\end{table}

For inferential purposes, the posterior mean vector and both joint and marginal intervals can be derived from
the Monte Carlo samples. However, as discussed below in Section \ref{Conclusion}, it may be useful to compute
the joint posterior mode for use as an estimator. For example, the posterior mode reduces to the MLE when the
parameter $\alpha_i =1$ for all $i$. A second advantage of this model is that the joint mean can be
computed through the Lindley-Smith maximization method  \citep{Lindley72,Chen2000}.

Lindley and Smith
proved that an algorithm that iteratively maximized each of the full conditional distributions will eventually
converge to the global maximum of the posterior. Because each of the conditionals in this model has a closed
form expression for its maximum, this optimization approach is both simple and extremely fast.  Similar to Gibbs sampling, this procedure starts from an arbitrary point.  Maximization proceeds by iterating
across the full set of conditional distributions, maximizing each and substituting the maximum into the next
conditional.  The hierarchical model considered here is not identical to eq.~(\ref{eq:mlike}), but depends
upon hyperparameters $(\gamma_1,\gamma_2)$ which effect the mode of the posterior distribution.  One possibility
for choosing these hyper-parameters is to pick them to minimize the distance between analytical mode and
iteratively computed mode.

\begin{algorithm}[!ht]
\SetLine
//\small{\textit{Initialization}}\\
Set the initial values $ \gamma_1, \gamma_2, \mathbf{N}, \mathbf{m}_0, \mathbf{g}$.

//\small{\textit{Main loop}}\\

\While{$\sum |m_i^{\mbox{new}}-m_i|>10^{-5}$}{
     $m_i^{\mbox{new}} = \max[\pi(m_i|t_i,g_i,N,\boldsymbol{\phi},\boldsymbol{\alpha},\gamma_1,\gamma_2] =
     (g_i+\alpha)/(N)$\\
     $g_i^{\mbox{new}} = \max[\pi(g_i|\mathbf{T},\mathbf{m},N,\boldsymbol{\phi},\boldsymbol{\alpha},\gamma_1,\gamma_2 ] = t_i +\mbox{floor}(Nm_i(1-\phi_i))$\\
     $N^{\mbox{new}} = \max[\pi(N|\mathbf{T},\mathbf{g},\mathbf{m},\boldsymbol{\phi},\boldsymbol{\alpha},\gamma_1,\gamma_2] = (\sum g_i+\gamma_1)/(1+\gamma_2)$}
\caption{Dirichlet-Poisson-Binomial posterior mode estimation via the Lindley-Smith algorithm} \label{algo1}
\end{algorithm}
\subsection{A Dirichlet-Multinomial-Binomial Approach.} \label{BinMulti}
%Gibbs2Unix1
%Gibbs2Unixn

A second approach is both more computationally efficient and arguably more directly comparable to the sampling
mechanism inherent in SAGE. For convenience, we begin by assuming that the total number of mRNA in the sample is
$N \sim Pois(\lambda)$. Given this mRNA population size, the vector $\mathbf{g}$ of counts of each category of
mRNA prior to tag formation follows a multinomial distribution,
\[[\mathbf{g}|\alpha] \sim {N \choose g_1, g_2, \ldots, g_k} m_1^{g_1} m_2^{g_2},\ldots, m_k^{g_k}\] whose probabilities are the $m_i$ and are assumed to follow a Dirichlet distribution with parameter vector $\mathbf{\alpha}$.

The joint posterior distribution can now be written as, \begin{eqnarray}
[\mathbf{m},\mathbf{g},N|\mathbf{T},\boldsymbol{\alpha},\gamma_1,\gamma_2] &\propto& \prod_{i=1}^{k} {g_i \choose t_i} \phi_i^{t_i}(1-\phi_i)^{g_i-t_i}
\frac{e^{-\lambda} \lambda^N}{N!} \nonumber\\
&& \times {N \choose g_1, g_2, \ldots, g_k} m_1^{g_1+\alpha_1-1} m_2^{g_2+\alpha_2-1},\ldots, m_k^{g_k+\alpha_k-1}. \nonumber \end{eqnarray}

Reminding readers that  $\boldsymbol{\phi}= (\phi_1, \phi_2, \ldots, \phi_k)$ be the vector of known gene dependent tagging probabilities, the full conditional distributions here are,
\begin{eqnarray}
\left[\mathbf{T}|t_i,g_i,N,\gamma_1,\gamma_2\right] & \sim & Dirichlet(\alpha + G) \nonumber \\
\left[\mathbf{g}|\mathbf{T},\mathbf{m},N,\boldsymbol{\alpha},\gamma_1,\gamma_2\right] & \sim & t_i + Multinom(N-T_t, \mathbf{m}*(1-\boldsymbol{\phi})) \nonumber \\
\left[N|\mathbf{T},\mathbf{g},\mathbf{m},\boldsymbol{\alpha},\gamma_1,\gamma_2\right] & \sim &
\Gamma(T_{tot}+\gamma_2,1+\gamma_1) \nonumber
\end{eqnarray}
where $\mathbf{m}*(1-\boldsymbol{\phi})$ is an element wise product with $1$ representing a vector of identical
dimension to $\mathbf{m}$. Hence, the $i^{th}$ element of the vector is given by $m_i(1-\phi_i)$.

We refer to this hierarchical model as the Dirichlet-Multinomial-Binomial (DMB) model.  Inference in this case is
much faster because the entire $\mathbf{m}$ vector can be updated at once instead of requiring separate updates
for each gene level $i$. This formulation is also more natural in the sense that it restricts the pre-tagging
counts to sum to the population total $N$. One drawback is that the observed data provides no information
for posterior inference about the population size $N$. A minor drawback of this approach is that it is more
difficult to perform a Lindley - Smith maximization to derive the mode. Maximizing the conditional distribution
of $\mathbf{g}$, the vector of mRNA counts, would require maximizing a multinomial distribution over the
discrete counts.

Experimental testing of this algorithm provided results similar to the first procedure. Five-million variates
were drawn with every hundredth stored.  The DMB algorithm was dramatically faster than the DPB approach
finishing simulation in just over 30 minutes.  Table \ref{DPBcor}
shows little if any correlation at any lag. For sample size $N$, this is to be expected since the conditional
distributions depend only on the data and hyperparameters which are fixed.  We conclude that the algorithm could
be accelerated without loss of accuracy by reducing the number of simulations. Inference was based on the final
1000 stored samples. Figure \ref{fig3} below displays inferential results from this model for the 20 most
frequently tagged genes ranked from largest to smallest. Behavior and trends are almost identical to the DPB
model.  For the case where all $\alpha_i = 1$, the posterior mean was systematically smaller than the
posterior-mode/MLE for cases with large tag counts. The posterior-mode was distinctly different than the
unadjusted tag proportions and both sets were above the upper credible bounds for the posterior mean.  In the
case where all $\alpha_i = 1/k$, the posterior-mean again matches the corrected MLE almost exactly. The
posterior mode tracks the mean closely but is always larger and the unadjusted proportions are usually outside
the credible region.

\begin{figure}[htbp]
\begin{center}
\caption{\label{fig3} Probability estimates for the Dirichlet -Multinomial -Binomial where all $\alpha_i=1$. The
20 genes with the largest tag counts are arranged in decreasing rank order along the X-axis. The observed tag
proportions are marked in dark circles, the standard MLE in dark triangles. The analytically computed posterior
mode $\alpha_i=1$ coincides exactly with the MLE.  Also included are the estimated posterior mean and upper and
lower 95\% Bayesian Credible bounds based on MCMC sampling.}
\includegraphics[scale=.75]{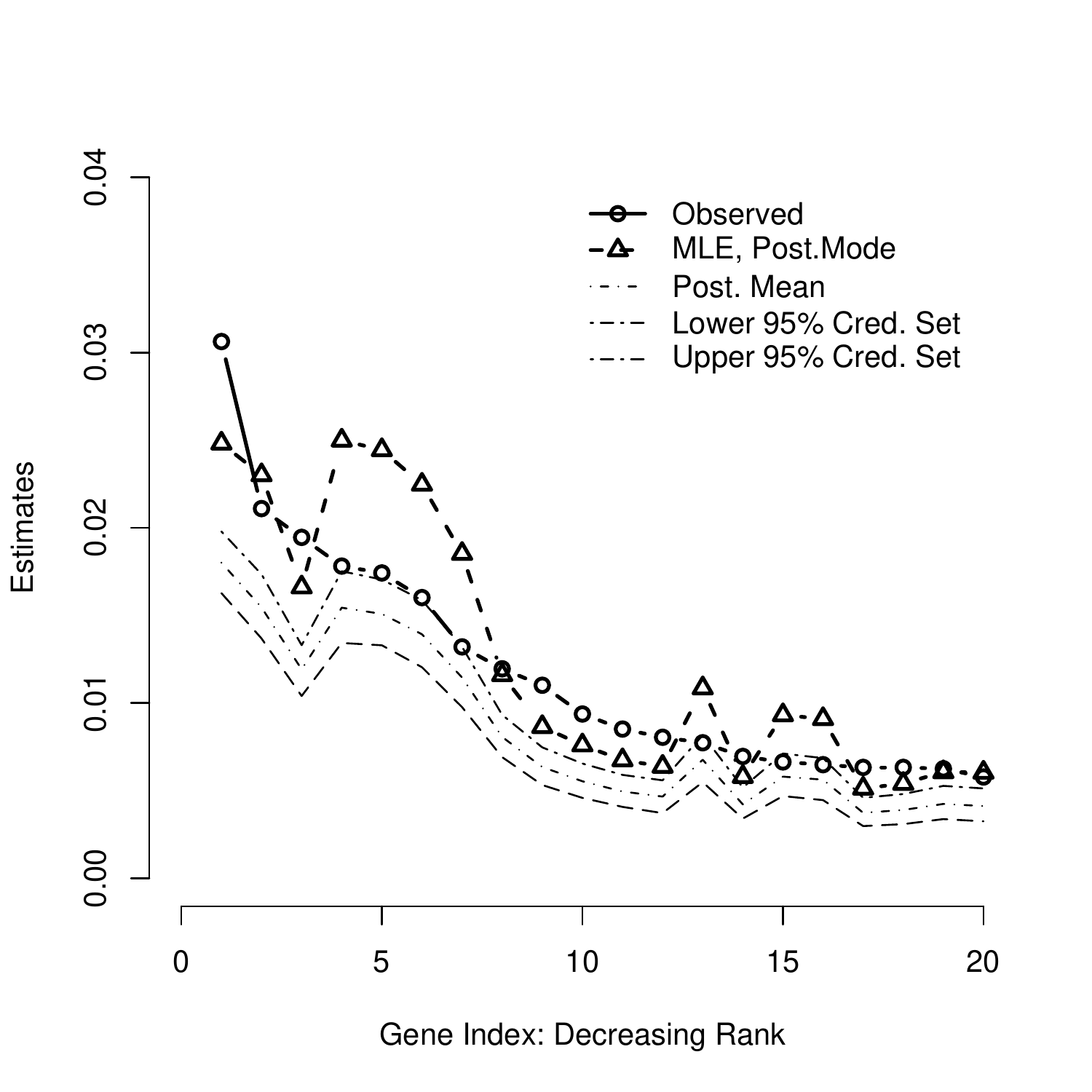}
\end{center}
\end{figure}

\begin{figure}[htbp]
\begin{center}
\caption{\label{fig4} Probability estimates for the Dirichlet -Multinomial -Binomial model with all
$\alpha_i=1/k$. The 20 genes with the largest tag counts are arranged in decreasing rank order along the X-axis.
The observed tag proportions are marked in dark circles, the standard MLE in dark triangles. In this case
analytically derived posterior modes deviate substantially from MLE.  However, estimated posterior mean is
identical to MLE in this case.  Upper and lower 95\% Bayesian Credible bounds are also given.}
\includegraphics[scale=.75]{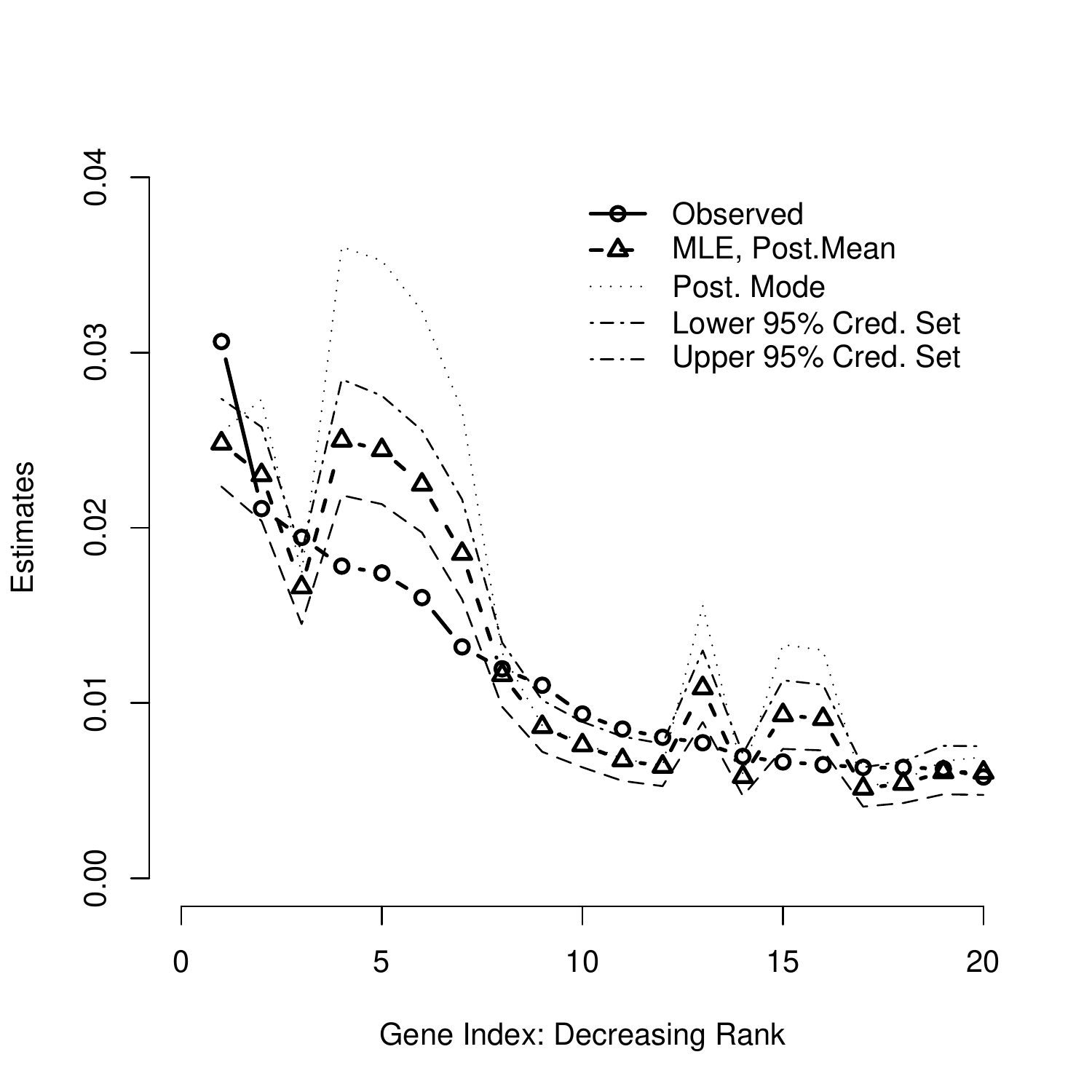}
\end{center}
\end{figure}

\subsection{A Missing Data Approach.\label{Conjugate}}

Instead of renormalizing the probabilities $\theta_i = (m_i \phi_i)/(\sum{m_i \phi_i})$ and doing inference
based on the posterior of eq.~(\ref{eq:postdist}), a second more straightforward approach uses missing
data.

Considering again a data generation process as described in section \ref{model}, we augment the observed data
vector $T$ with an extra category which represents any cDNA that is not converted to a SAGE tag. This count is
not observed, but if a distribution such as a Poisson is proposed, we may consider a Bayesian approach.  The
resulting posterior distribution for the data is,
\[[\mathbf{m}|\mathbf{T},\boldsymbol{\alpha}] \propto (1-\sum_{i=1}^k m_i \phi_i)^r
\prod_{i=1}^k (m_i \phi_i)^{t_i+\alpha_i-1} \]
%{T_t+r \choose t_1,t_2, \ldots, t_k,r}
where we have assumed a Dirichlet prior on the unknown initial proportions $\mathbf{m} \sim
\mathrm{D}_k(\alpha_1, \ldots, \alpha_k)$ . The resulting posterior distribution is a Dirichlet
distribution with $k+1$ categories, compared with the $k$ categories in the initial distribution $\mathbf{m}$. In
particular the distribution is a $\mathrm{D}_{k+1}(t_1+\alpha_1, \ldots,t_k+\alpha_k,r+1)$.

By integrating over the missing data term or otherwise eliminating it, it is possible to sample from the joint
distribution whose success probability is $m_i \phi_i$. According to the marginalization property of the
Dirichlet \citep{Bilodeau1999} the distribution of any subset of variables $(\theta_{i_1}, \theta_{i_2}, \ldots,
\theta_{i_j})$ from a $\mathrm{D}_k(\alpha_1,\alpha_2,\ldots,\alpha_k)$  with $\sum \alpha_ i = p$ is
distributed $\mathrm{D}_{j+1}(\alpha_{i_1}, \alpha_{i_2}, \ldots, \alpha_{i_j},q)$. Here $q = p-\sum_{i=1}^{j}
\alpha_{i_j}$ corresponds to the total of the variables being marginalized over. In words, by considering only a
subset of j categories we observe a Dirichlet distribution on j+1 categories with $\sum_i \alpha_i=p$ remaining
the same. In our case, the value of $r$ will be a component of the parameter $q$.  Several modes of inference are available based on this missing data approach. Suppose first that we are
particularly interested in the posterior means of the $m_i$ parameters.   An exact calculation is possible based
upon the marginalization properties of the Dirichlet \cite[Corollary 3.4]{Bilodeau1999}.
\begin{thm}
Let $\boldsymbol{\theta} = (\theta_1, \ldots, \theta_p, \theta_{p+1}) \sim \mathrm{Dirichlet}(\alpha_1, \ldots,
\alpha_p, \alpha_{p+1})$.  Let $\alpha_0 = \sum_{i=1}^{p+1}{\alpha_i}$.  Suppose that $S= \sum_{i=1}^p
\theta_i$, and for $j<p$, let $\boldsymbol{\theta}_1 =( \theta_{i1}, \ldots, \theta_{ij})$ be any subset of
components of $\boldsymbol{\theta}$.  It follows that $\boldsymbol{\delta_1} = {\theta}_1/S \sim
\mathrm{Dirichlet}(\alpha_i1, \ldots, \alpha_ij,\delta)$ with $\alpha_0 - \alpha_{p+1} = \delta + \sum_{d=1}^j
\alpha_{id}$
\end{thm}
This result demonstrates that, ignoring the missing categories, the distribution of the
\[\frac{(m_1 \phi_1, \ldots, m_{k-1} \phi_{k-1}, m_k \phi_k)}{\sum_1^k m_i \phi_i} \sim \mathrm{Dirichlet}(\alpha_1+t_1, \ldots, \alpha_{k-1}+t_{k-1},\delta), \]
with $\delta = t_k + \alpha_k$.  Let $w_i = (\alpha_i + t_i)/\sum(\alpha_i + t_i)$. Then by reweighting, we can
compute,
\[E(m_i) = \frac{1}{\phi_i \sum_i w_i}\frac{\alpha_i + t_i}{\sum (\alpha_i + t_i)}. \]

More generally, Gibbs sampling is also available for this problem and extends inference to Bayesian interval
estimates. Again observing the posterior distribution,

\begin{equation}
[\mathbf{m},r|\mathbf{T},\mu] \propto (1-\sum_{i=1}^k m_i \phi_i)^r \prod_{i=1}^k
(m_i\phi_i)^{t_i+\alpha_i-1} \frac{e^{-\mu} \mu^r}{r!}
\end{equation} The full conditional distributions are derived to be,
\begin{eqnarray*}
\left[r|\mathbf{T},\mathbf{m},\boldsymbol{\alpha},\boldsymbol{\mu}\right] & = & Poisson((1-\sum m_i \phi_i) \mu) \\
\left[\mathbf{m}\boldsymbol{\phi}|\mathbf{T},r,\boldsymbol{\phi},\boldsymbol{\alpha},\boldsymbol{\mu}\right] & = & D(r,\alpha_1+t_1, \ldots, \alpha_k+t_k)  \\
\end{eqnarray*}
In order to apply Gibbs Sampler a value for the hyperparameter $\mu$ must be selected.  In the case where all
$\alpha_i=1$, a logical mean for the Poisson is $T_{tot}\sum_i m_i(1-\phi_i)/\sum_i m_i \phi_i$ since
$\sum m_i (1-\phi_i) = 1 - \sum m_i \phi_i$ is the probability that an mRNA is not converted into a tag.
Equation \ref{eq:implicit.mhat} provides a useful estimate of $\sum m_i\phi_i$.

\begin{figure}[htbp] \caption{\label{fig5} Probability estimates for the Missing Data Algorithm where all $\alpha_i=1$.
The 20 genes with the largest tag counts are arranged in decreasing rank order along the X-axis. The observed
tag proportions are marked in dark circles, the standard MLE in dark triangles. The analytically computed
posterior mode $\alpha_i=1$ coincides exactly with the MLE.  Also included are the estimated posterior mean and
upper and lower 95\% Bayesian Credible bounds based on MCMC sampling.}
\begin{center}
\includegraphics[scale=.75]{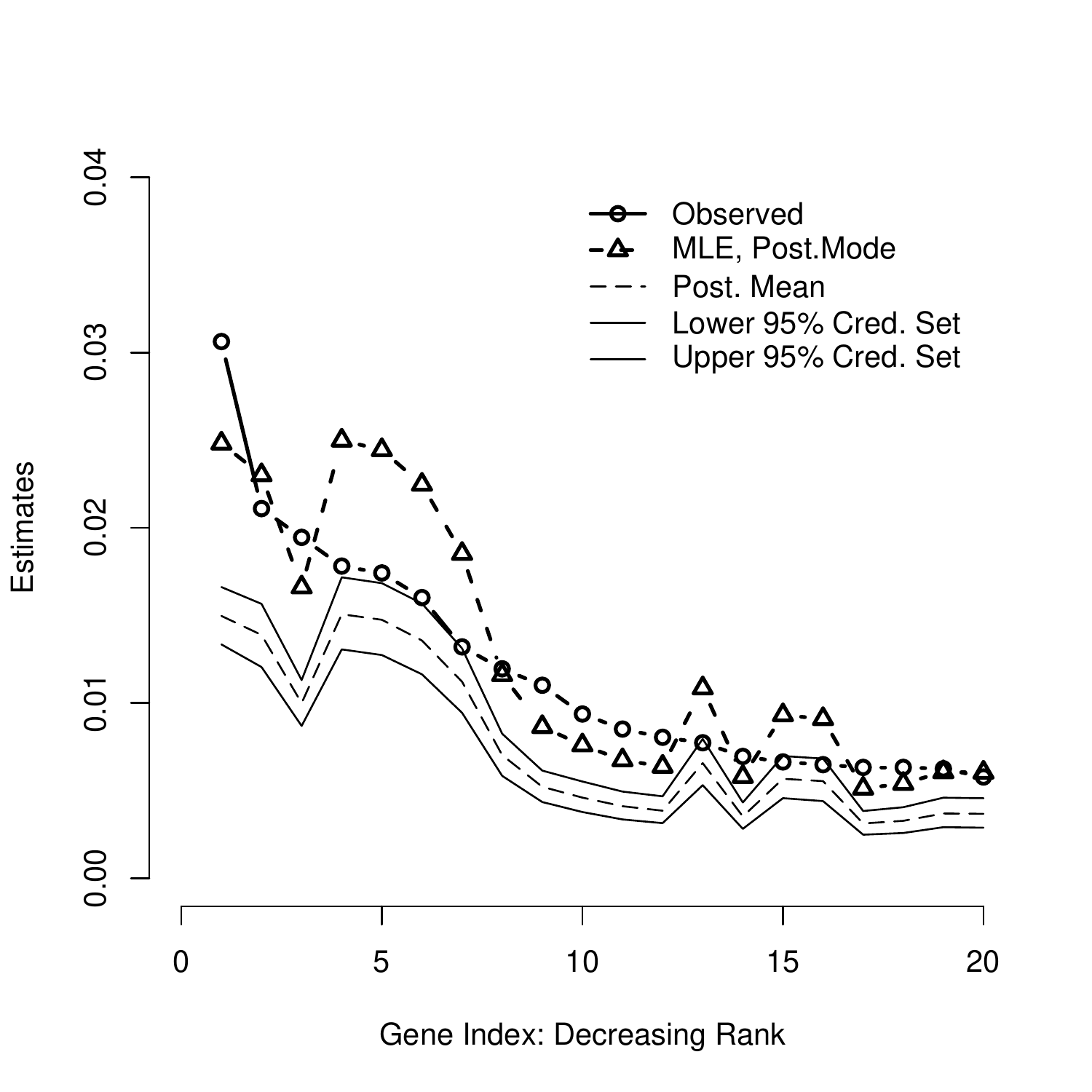}
\end{center}
\end{figure}

\begin{figure}[htbp]
\begin{center}
\caption{\label{fig6} Probability estimates for the Missing Data Algorithm where $\alpha_i=1/k$. The 20 genes
with the largest tag counts are arranged in decreasing rank order along the X-axis. The observed tag proportions
are marked in dark circles, the standard MLE in dark triangles. In this case analytically derived posterior
modes deviate substantially from MLE. However, estimated posterior mean is identical to MLE in this case.  Upper
and lower 95\% Bayesian Credible bounds are also given. }
\includegraphics[scale=.75]{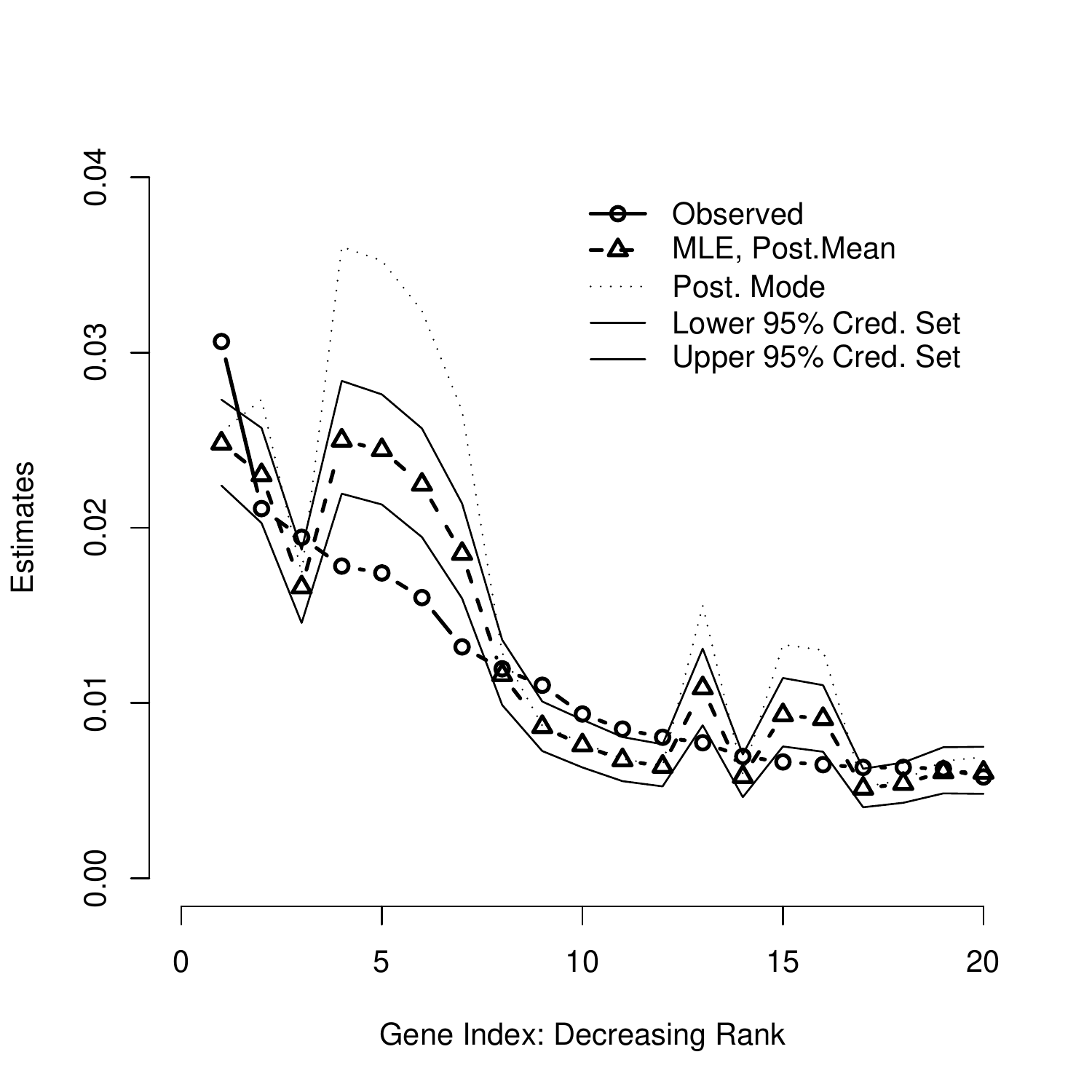}
\end{center}
\end{figure}

Like the DPB algorithm above, the missing data algorithm also admits a simple Lindley-Smith optimization procedure in order to compute the posterior mode.  If the prior mean of $r$, $\boldsymbol{\mu}=10,100$ then the
posterior mode of the missing data model is nearly identical to the exact analytical mode. The mode of the Dirichlet conditional is
\begin{equation}
\widehat{m_i \phi_i} = \frac{t_i + \alpha_i - 1}{\sum_{i} (\alpha_i + t_i) + r - k}, \end{equation} see
\cite{Gelman2005}. The mode for the Poisson is the mean rounded down to the nearest smaller integer \citep{Wiki:Poisson}.  Results for this case are very similar to the previous two.  This algorithm is
clearly the fastest, completing 5 million samples in  20 minutes. Autocorrelation
in the untagged population size $r$ decreases to 0 after 50 simulations but proportions $m_i$ seem uncorrelated
at all lags.  Conclusions for Figures \ref{fig5} and \ref{fig6} are largely identical to the above.  In fact, as
implemented quantitative differences do exist between the methods in the case where all $\alpha_i = 1$ with
this method providing the smallest or most heavily shrunk estimates of genes with large counts and the DPB
method giving the largest estimates. Because probabilities must sum to one, the opposite effect should exist for
categories with small counts although the mass will be spread over thousands of categories leading to very
minute differences.

\section{Conclusions}\label{Conclusion}
Sampling bias is a ubiquitous problem in studying biological problems.  This work focuses on a general biasing mechanism of which data generation from in SAGE is a protype. We present
three hierarchical models which provide posterior inference for the biased sampling model given in eq.
(\ref{eq:mlike}).  This work compliments earlier work by \citet{Gilchrist2007} in two notable ways. First, it
provides a flexible and efficient methodology which extends the range of the prior distributions that are
available.  It is important to point out the effect of the typical flat prior $\boldsymbol{\alpha}=1$ on
inference: analogously to the binomial case \citep{Christensen1997}, this prior assumes that each category was
seen once in previous experiments, a very strong assumption due to the small size of the tag sample compared to
the number of categories.  The result is that the prior overwhelms the data for many of the categories.  In the
context of SAGE, it may be more appealing to the scientist to choose a weaker prior such as $\boldsymbol{\alpha}
= 1/k$, which, as we've seen, gives the actual data more voice. Because the analytical approximations of
\cite{Gilchrist2007} do not extend easily to this regime of priors, the simulation based results presented here
are increasingly valuable. Second, the more robust methodology presented here provides a foundation for extending the analysis to include more realistic models of the tag formation process (such as the inclusion of transcription or sequencing errors) or more complex priors.

Our approach may be contrasted with \cite{Morris2003}, who investigated joint estimators of the relative
proportions of different mRNA transcripts in SAGE experiments based on posterior inference from a conjugate
Dirichlet-Multinomial model. They note that while the shrinkage properties of the
posterior mean provide improved average inference with respect to squared error loss across all categories, they
tend to shrink categories where large counts are observed excessively in order to boost the estimated
probabilities of cells with few or no counts.  Due to the massive number of categories and the extreme imbalance
in observed frequencies (e.g.~more than half of the categories have zero counts in our data), the estimates for
frequently observed genes tend to be fairly poor.  This observation highlights the main weakness of shrinkage
estimators, while they perform better on average across categories, they may perform poorly on particular
categories which may be of primary interest. \cite{Morris2003} resolve the problem by introducing
a hierarchical Dirichlet-Multinomial model which clusters genes into two categories, high and low frequency,
partitioning the overall mass in order to mitigate the shrinkage in the high frequency categories.

Our contribution is clearly distinct from that of \cite{Morris2003} since
it focuses on a discrepancy in the sampling model.
Nevertheless, it should be possible to extend our methods to allow for the mixture priors used by \cite{Morris2003}.
Even without such an extension, our results offer alternative methods and important
insights into shrinkage in models with large or massive numbers of categories.  For example, we've shown
that by using the prior $\boldsymbol{\alpha} = 1/k$, the posterior mean is equal to the corrected MLE.  The
additional advantage of a sampling model is that it generates marginal posterior credible sets for the
proportion $m_i$. If inference focuses only on genes with large expression levels, we gain one of the main
advantages of Monte Carlo sampling without the concern of excessive shrinkage.  If one is only interested in
point estimates, and excessive shrinkage is a concern, it may also be appropriate to use the analytical mode as an estimator.  For example, the posterior mode for model \ref{eq:postdist} is identical to the MLE for model \ref{eq:mlike} when $\boldsymbol{\alpha} = 1$ , while the
posterior mean becomes equivalent when $\boldsymbol{\alpha} = 1/k$.  Given the stated hyper-parameters,
posterior means are shrunk for more frequently observed categories and increased for rarely observed categories.
The models provide different levels of shrinkage, with the Dirichlet-Poisson-Binomial (DPB) model providing the least,  and
the missing data model providing the most shrinkage.  In terms of computation time, the DPB method is much less
efficient and arguably offers no advantage over the other methods for SAGE.  However, in  other applications the
DPB model may be closer to the model of interest.

A number of important statistical questions remain to be answered.  Besides the sampling
errors dealt with here, there are other sources of errors inherent in SAGE's tag formation and sampling processes
\citep{Stollberg2000}, which future work could address. This analysis also presumed that the category dependent sampling probabilities were fixed when in fact
they were estimated from the data. A model that views $\phi$ as a random quantity would also represent an
improvement. Finally, a major emphasis of SAGE analysis regards evaluation of differential expression levels
\citep{Baggerly2003,Baggerly2004} between related cell samples.  Because the AE cutting probability $p$ is
experiment dependent \citep{Gilchrist2007}, it is important that the biasing mechanism be taken into account when
evaluating differential expression across experiments. Indeed, this should lead to increased power and accuracy
in such studies.
\bibliographystyle{apacite}
\bibliography{Sage2}

\end{document}